\documentclass[conference]{IEEEtran}
\IEEEoverridecommandlockouts

\usepackage{cite}
\usepackage{amsmath,amssymb,amsfonts}
\usepackage{algorithm}
\usepackage{algorithmic}
\usepackage{graphicx}
\usepackage{textcomp}
\usepackage{xcolor}
\usepackage{xspace}
\usepackage{soul}
\newcommand{\BfPara}[1]{\vspace{1mm}{\noindent\bf#1.}\xspace}

\def\BibTeX{{\rm B\kern-.05em{\sc i\kern-.025em b}\kern-.08em
    T\kern-.1667em\lower.7ex\hbox{E}\kern-.125emX}}
\begin{document}

\title{Reinforced Edge Selection using Deep Learning for Robust Surveillance in Unmanned Aerial Vehicles}

\author{\IEEEauthorblockN{$^{\circ}$Soohyun Park, $^{\dag}$Jeman Park, $^{\S}$David Mohaisen, and $^{\circ,\ddag}$Joongheon Kim}
\IEEEauthorblockA{$^{\circ}$School of Electrical Engineering, Korea University, Seoul, Republic of Korea 
\\$^{\dag}$School of Electrical and Computer Engineering, Georgia Institute of Technology, Atlanta, GA, USA
\\$^{\S}$Department of Computer Science, University of Central Florida, Orlando, FL, USA
\\$^{\ddag}$Artificial Intelligence Engineering Research Center, College of Engineering, Korea University, Seoul, Republic of Korea
\\
E-mails: \texttt{soohyun828@korea.ac.kr}, 
\texttt{parkjeman122@gmail.com},\\
\texttt{mohaisen@ucf.edu},
\texttt{joongheon@korea.ac.kr}
}
}
\maketitle

\begin{abstract}
In this paper, we propose a novel deep Q-network (DQN)-based edge selection algorithm designed specifically for real-time surveillance in unmanned aerial vehicle (UAV) networks. The proposed algorithm is designed under the consideration of delay, energy, and overflow as optimizations to ensure real-time properties while striking a balance for other environment-related parameters. The merit of the proposed algorithm is verified via simulation-based performance evaluation. 
\end{abstract}

\section{Introduction}
In large-scale industrial environments such as smart factories and smart harbors, it is necessary to detect and analyze unpredictable and anomalous situations~\cite{IEEE Access,MDPI1,APNOMS}. Given that those environments are large-scale by design, employing traditional surveillance systems that require significant infrastructure investment is impractical. 
For this purpose, the use of unmanned aerial vehicles (UAVs) is one of the more promising solutions to provide surveillance capabilities while addressing a flexible network infrastructure deployment. However, because UAVs have various limitations in terms of energy (i.e., battery) capacity, the UAVs for surveillance applications require various optimizations in terms of energy-efficient methods for flight time extension, efficient mobile computing, and fast and scalable storage management, among others.
In this paper, the UAVs are used for surveillance purposes, thus, the energy-efficiency for recording video streams is also essential, which constitutes the main focus of this work.
It is noted that there have been several studies on using UAVs in smart factories or smart harbor environments for surveillance purposes, providing grounds for this work, where the main purpose of the research results for the UAV-based surveillance networks has been to ensure efficient monitoring for surrounding structures/facilities. The main goal of our work in this paper is to facilitate the deployment of such systems through various optimizations. 

\BfPara{Contribution} In this paper, we propose a system using UAVs that transmits camera-based observation data to the surrounding edges to overcome the limitations of UAVs and also to ensure robust management of the monitoring systems in real-time. To ensure real-time optimizations and extended operation of the system, we further propose a deep reinforcement learning-based decision algorithm that we use to decide which edges will receive the surveillance data from the UAV. The proposed edge selection algorithm is trained to perform the best action (optimal action) for maximizing reward where the reward is organized under the consideration of delay, energy, and overflow (stability).

\BfPara{Oranization} The rest of this paper is organized as follows
Sec.~\ref{sec:sec2} introduces reference network models in this paper. 
Sec.~\ref{sec:sec3} presents the details of the proposed deep reinforcement learning based edge selection for robust and real-time surveillance in UAV networks. 
Sec.~\ref{sec:sec4} evaluates the performance of the proposed algorithm.
Lastly, Sec.~\ref{sec:sec5} concludes this paper and then presents future work directions.

\begin{figure}
    \centering
    \includegraphics[width=1.0\columnwidth]{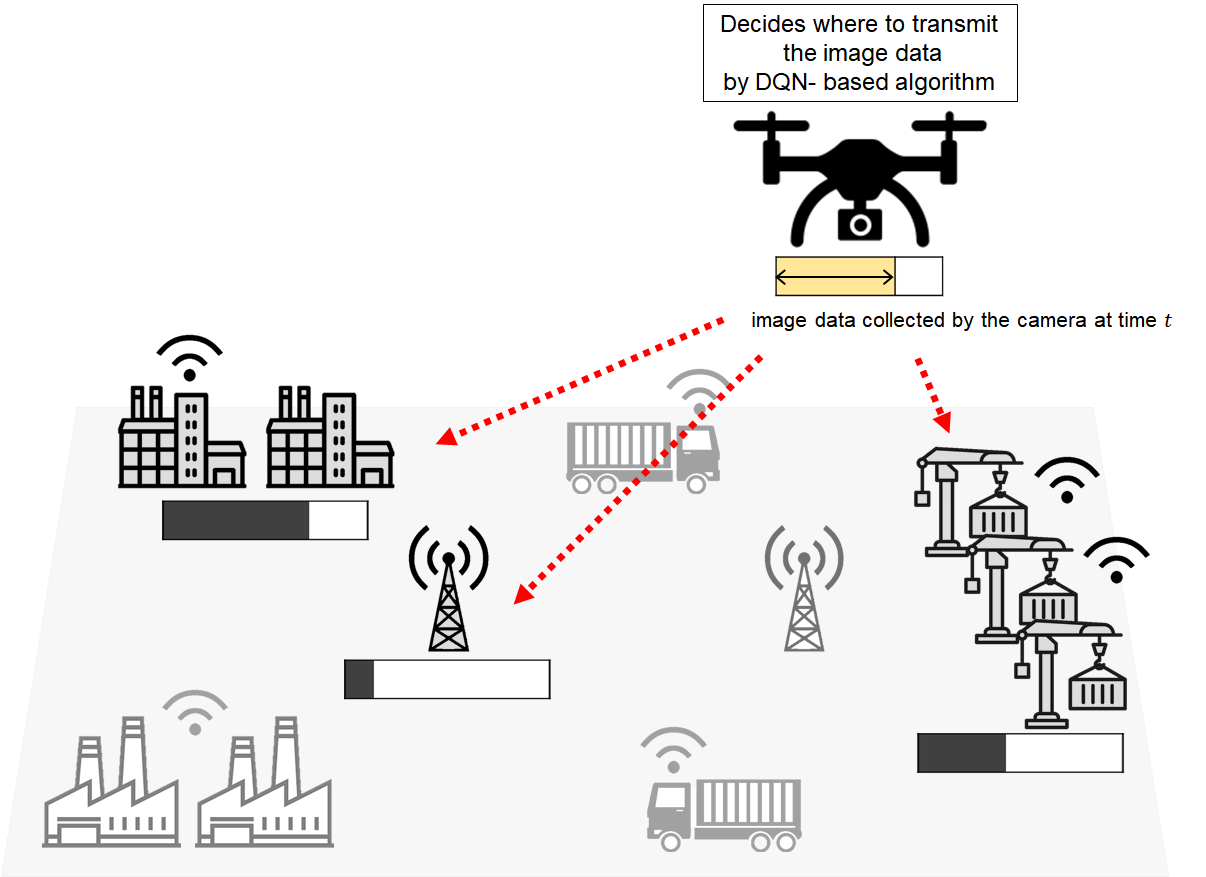}
    \caption{Reference network model.}
    \label{fig:fig1}
\end{figure}

\section{Reference Network Model}\label{sec:sec2}
In this paper, we consider a reference network model that can automatically gather and analyze surveillance images from CCTV-camera-equipped UAVs. The overall architecture of our system is as illustrated in Fig.~\ref{fig:fig1}. In this system, the UAVs continuously observe the conditions of the environment and collect image data through a built-in camera by flying over the smart factory or smart harbor environments. 
The collected surveillance data can be used for detecting or predicting multiple risks, anomalies, or problems. 
In this process, it can be more efficient to transmit the images to the edge which is located nearby and also has more computing power and capabilities that would allow it to process images quickly and meet the application's deadlines. 
The main purpose of the proposed algorithm is to extend the UAV's flight time and improving the quality of image-based real-time management and accident prediction system module and objective. 

For this purpose, we design an algorithm using reinforcement learning. In this paper, the proposed algorithm is designed based on reinforcement learning because it allows us to sequentially control the actions/decisions in the system depending on time-varying states/environments for maximizing our rewards/objects. The proposed algorithm finally utilizes deep Q-network (DQN) which is designed for solving reinforcement learning using deep neural networks (also called deep learning) for fast and approximated large scale problem solving in many applications as demonstrated in the literature~\cite{rl1,rl2,rl3,rl4,rl5,rl6,rl7,rl8,rl9,rl10,rl11,rl12,rl13,rl14}.

\section{DQN-based Edge Selection}\label{sec:sec3}
As mentioned above, we use DQN, one of the well-studied reinforcement learning methods in the literature, for our application. DQN is trained by deep neural network frameworks for the proposed edge selection algorithm. Moreover, DQN is designed so as to allow an agent to behave in order to receive the largest reward at all moments using the separated target network and a neural network learning process through multiple layers. This section introduces the proposed DQN-based edge selection algorithm. 

In the proposed DQN-based edge selection algorithm, states, actions, and rewards are defined as follows:

\begin{itemize}
    \item \textbf{State:} 2-types of observations are considered as states, i.e., the information of the agent (UAV backlog, input image data size, data size for transmission, and energy) and information of the environment (edge capacity, and the distance between an edge and its corresponding agent). The data about the environment contains not only one edge information, but also information associated with multi-edge in the system.
    \item \textbf{Action:} The action space size in our algorithm is equivalent to the number of edges in the system. The agent can choose one integer value between $0$ and $N$ as the number of edges, and transmit the data to the selected number of edges. The next state and current reward will be determined by this current action.
    \item \textbf{Reward:} The agent receives reward for the action upon taking such action. The reinforcement learning algorithm aims to learn the optimal actions and get the optimal policy for the maximum rewards at the end. In this paper, the following aspects are considered as rewards: (i) delay before the transmitted data are processed ($R_{d}$ below), (ii) energy consumption for flight and transmission ($R_{e}$ below), and (iii) overflow occurred at the selected edge ($R_{o}$ below, negative reward). These rewards are considered when the agent takes actions, and they can be formulated as follows:
\begin{equation}
R = \alpha R_{d} + \beta R_{e} + (1-(\alpha + \beta)) R_{o}
\label{eq:reward}
\end{equation}
where $R_{d}$ and $R_{e}$ are always positive for the actions of UAV, they have smaller values when the data transmission to edges occurs. The $R_{o}$ is only calculated when the edge overflow is originated, and this decreases the total reward. The agent considers all the above aspects to obtain maximum reward.
\end{itemize}

\begin{figure}[!t]
    \centering
    \includegraphics[width=1\columnwidth]{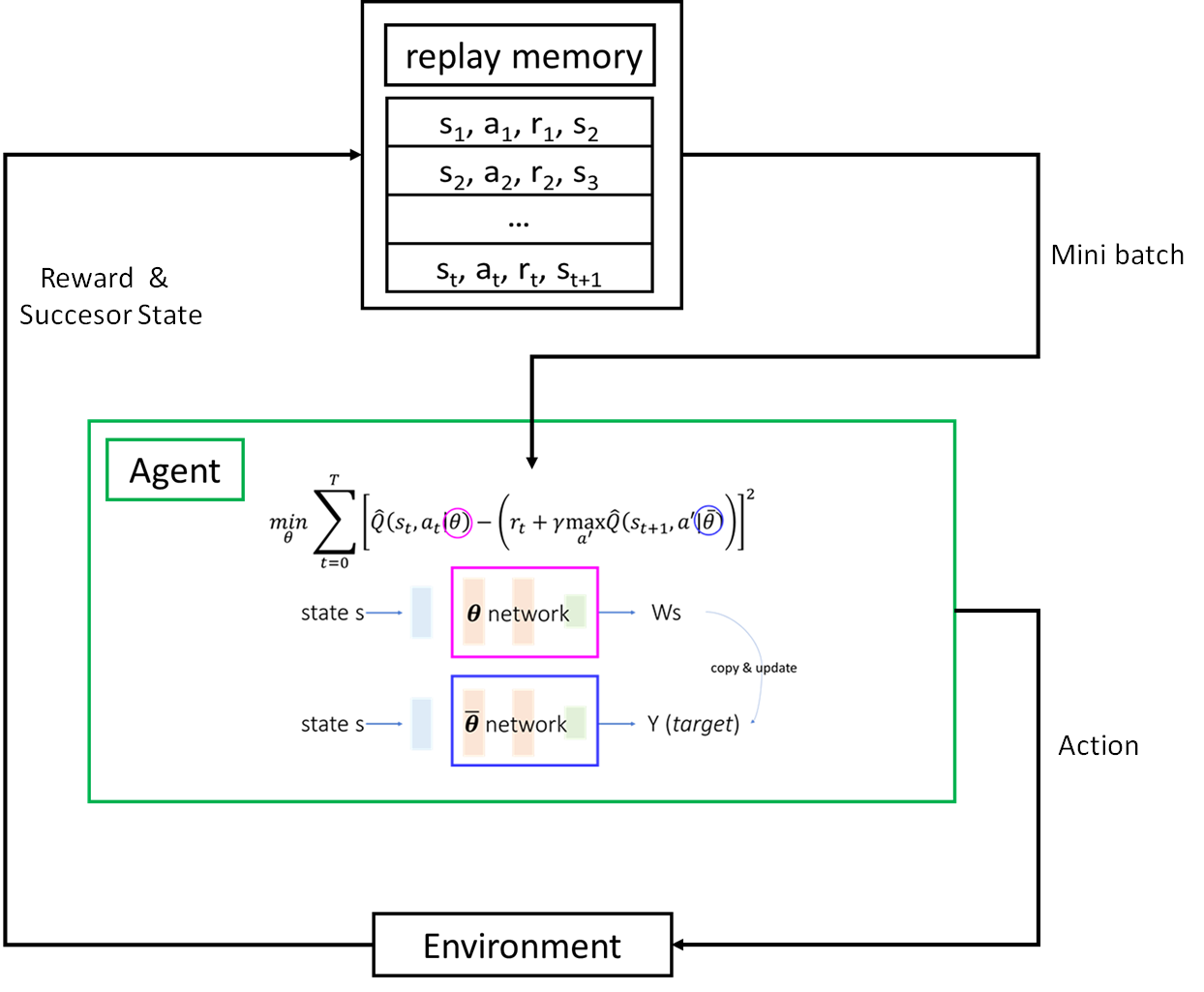}
    \caption{Diagram of the DQN algorithm.}
    \label{fig:fig2}
\end{figure}

The proposed edge selection algorithm is executed as illustraetd inFig.~\ref{fig:fig2}. The DQN agent is trained using the reply memory and the separated target network. The two components are the main components for DQN computation. According to the theory of reinforcement learning, the agent's sequential action making which is transmitted to the environment is effected to the next time-step state and the reward. The information of state and reward is stored in the replay memory. When the agent takes actions, the corresponding rewards are calculated by gradient descent optimization steps with respect to the network parameter $\theta$. The target network parameter is reset by $\theta$ at certain intervals (off-policy training).

\section{Performance Evaluation}\label{sec:sec4}

In this section, we describe the simulation environment and results to show the merits and performance of the proposed DQN-based algorithm. We assume that there exist a single UAV and 10 edges in the system. When every episode begins, the UAV and edges have randomly allocated backlogs that are limited by their own maximum capacities. 
In each time step, the input data size where the input is obtained by the camera and the output data size which will be transmitted to the edge are randomly selected between $x$ and $y$ or $w$ and $z$.
The values that we used in this simulation are in Table~\ref{tab1}.
If the agent (UAV) runs out of energy, the simulation is terminated. 
In our DQN learning, the input size of the model is defined by the number of edges in the system. The observation includes the number of edges of information about the agent and environment (i.e., UAV backlogs, input data size, output data size, UAV energy, the number of edges, edge capacity, and distances between edge and UAV).
Similarly, the output size is the number of dimensions of actions (where each edge will be selected). In order to train the model, the corresponding hyper-parameters are chosen as follows.
\begin{itemize}
    \item Discount rate: $0.99$
    \item Replay memory: $50,000$
    \item Batch size: $64$
    \item Target update frequency: $5$
    \item Total episodes: $5,000$
\end{itemize}

\begin{table}[t]
\caption{Simulation parameters}
\begin{center}
\begin{tabular}{|l|r|}
\hline
\textbf{Parameters}&{\textbf{Values}} \\
\hline
UAV capacity  & $250$ (GB)\\
Edge capacity & $1,500$ (GB)\\
Distance & min: $1$, max: $15$ (m)\\
UAV data rate & $0.9$ (GHz)\\
Edge clock frequency & $2.7$ (GHz)\\
Power consumption of transmission & $0.4$ / $0.7$ (power efficiency $70$\%)\\
Power consumption of moving & $0.6$ / $0.7$ (power efficiency $70$\%)\\
\hline
\end{tabular}
\label{tab1}
\end{center}
\end{table}

\begin{figure}[!t]
    \centering
    \includegraphics[width=1\columnwidth]{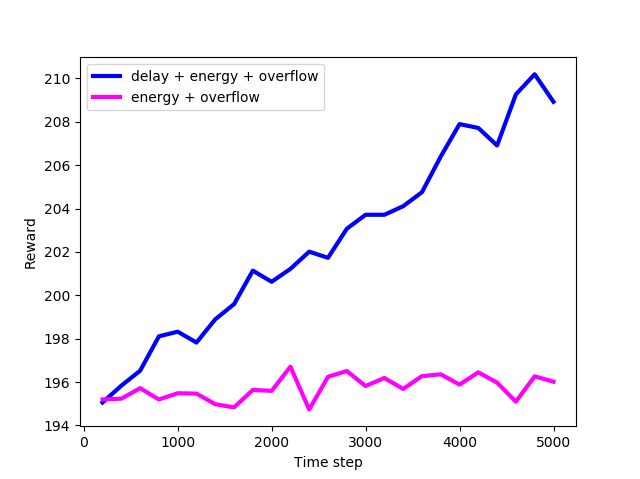}
    \caption{Accumulated Rewards in each time step.}
    \label{fig:fig3}
\end{figure}

Fig.~\ref{fig:fig3} shows our experimental results in terms of rewards where the reward equation is formulated as in (\ref{eq:reward}).
In order to compare our proposed algorithm with others, another reward formulation using two aspects (i.e., energy consumption and edge overflow) is used for comparison and as a baseline.  

Fig.~\ref{fig:fig3} represents the learning rewards until the end of the total number of episodes used in this study. 
The blue line represents the values of the basic reward equation in (\ref{eq:reward}) and the pink line stands for the values of the transformed reward equation under the consideration of energy consumption and edge overflow. For the basic reward in (\ref{eq:reward}), our proposed algorithm achieves more reward values as the time (i.e., episode) goes by. 
During the total episodes used in this experiment, the basic reward has a maximum value of $210$, which happens between episode $4,500$ and $5,000$, compared to the beginning. On the other hand, the transformed reward under the consideration of energy consumption and edge overflow reaches the biggest value between $196$ and $197$ and it does not increase over that value until the end of episode timeline. 

\section{Conclusions and Future Work}\label{sec:sec5}
This paper proposes a reinforcement learning-based edge selection algorithm designed for real-time surveillance in unmanned aerial vehicle (UAV) networks. 
In order to compute the reinforcement learning-based algorithm when the given states are large, deep learning-based computation (i.e., deep Q-network) is used for approximated reinforcement learning computation. 
The proposed DQN-based edge selection algorithm is designed under the consideration of delay, energy, and overflow. The merit of this algorithm is verified via simulation-based performance evaluation.

\section*{Acknowledgment}
This research was supported by Information \& Communications Technology Promotion (IITP) grant funded by the Korea government (MSIT) (No.2018-0-00170, Virtual Presence in Moving Objects through 5G). All authors in this paper have equal contributions (first authors). Joongheon Kim and David Mohaisen are corresponding authors.

\end{document}